\documentclass[12pt,a4paper]{article}

\usepackage[english]{babel}
\usepackage[T2A]{fontenc}
\usepackage[cp1251]{inputenc}
\usepackage{amsmath}
\usepackage{graphicx}
\usepackage{amssymb}
\usepackage{color}
\usepackage{amsfonts}
\usepackage{wrapfig}
\usepackage{caption}

\usepackage{hyperref}
\usepackage{physics}

\begin{document}
	
	\binoppenalty=10000
	\relpenalty=10000
	
\begin{center}
	\textbf{\Large{Thermodynamic response functions in a cell fluid model}}
\end{center}

\vspace{0.3cm}

\begin{center}
 O.A.~Dobush\footnote{e-mail:  dobush@icmp.lviv.ua}, M.P.~Kozlovskii, R.V.~Romanik\footnote{e-mail:  romanik@icmp.lviv.ua}, I.V.~Pylyuk 
\end{center}

\begin{center}
	Institute for Condensed Matter Physics of the National Academy of Sciences of
	Ukraine \\ 1, Svientsitskii Street, 79011 Lviv, Ukraine
\end{center}

	 \vspace{0.2cm}

	\begin{center}
 \small \textbf{Abstract} 
	\end{center}
	
	\small Thermodynamic response functions, namely the isothermal compressibility, the thermal pressure coefficient, and the thermal expansion coefficient, are calculated for a many-particle system interacting through a modified Morse potential. These calculations are based on an equation of state previously derived for a cell fluid model in the grand canonical ensemble. The calculated quantities are presented graphically as functions of density and the effective chemical potential.

	\normalsize 
\section{Introduction}

Thermodynamic response functions play a crucial role in understanding and characterizing the behavior of physical systems. These functions describe how a system responds to changes in its state variables, providing valuable insights into its thermodynamic properties. The most widely studied thermodynamic response functions of fluids are isothermal and adiabatic compressibilities, isobaric thermal expansion, isochoric thermal pressure coefficient and heat capacities either at constant pressure or at constant volume. Thermodynamic response functions are essential tools for understanding, predicting, and controlling the behavior of physical systems. They are used in various scientific and engineering disciplines to model, design, and optimize processes and to gain insights into the fundamental principles governing thermodynamic systems. Therefore, a key area of research involves examining the thermodynamic properties of simple fluids and fluid models through theoretical approaches, computer simulations, and experimental studies, covering both subcritical and supercritical regions~\cite{Johnston2014book,YigzaweSadus2013,Velasco2011,Bulavin2024}.

In this work we continue our study of the thermodynamic behavior of a cell fluid model, which was defined in~\cite{KozitskyKozlovskiiDobush2018book,KozitskyKozlovskiiDobush2020}. This model was used with the Morse potential in studies~\cite{PylyukEtAlJML2023,PylyukKozlovskiiDobushUJP2023b} and with a modified Morse potential in~\cite{KozlovskiiDobush2020,PylyukDobush2020} as potentials describing the particle interaction. In particular, the equation of state was obtained in~\cite{KozlovskiiDobush2020} in the zero-mode approximation. This equation of state is used in the current work to calculate thermodynamic response functions, namely the isothermal compressibility, the thermal pressure coefficient, and the thermal expansion coefficient. 

As the problem has been considered in the framework of the grand canonical ensemble, the initial equation of state is formulated in terms of pressure $P$, temperature $T$ and the chemical potential $\mu$, $P = P(T,\mu)$. To leverage this form of the equation of state for calculation of the response functions, the corresponding definitions should be transformed to proper derivatives with respect to temperature and chemical potential~\cite{StrokerMeier2021}. For each response function considered in this paper, we present such transformation in a dedicated subsection. On the other hand, within the approach applied here, the chemical potential can be explicitly expressed via the number particle density $\rho$ and temperature. This gives rise to the equation of state in the form $P = P(T, \rho)$, in which case it is suitable to re-express the response functions in terms of derivatives with respect to temperature and density.

In Section~\ref{sec:potential}, we present the modified Morse potential used in our work, and briefly compare it with other possible modifications. Section~\ref{sec:eos} is dedicated to the explicit expressions for the equation of state. In Section~\ref{sec:respons_functions}, the expressions for response functions are derived in terms of thermodynamic derivatives suitable for different forms of the equation of state.

\section{\label{sec:potential}The interaction potential}
The potential of interaction between particles is taken in a form of a modified Morse potential
\begin{eqnarray}
	\label{def:mod_morse}
	U(r) & = & \varepsilon C_{H} \left[A {\rm e}^{-n_0(r-R_0)/\alpha}  +  {\rm e}^{-\gamma(r - R_0)/\alpha} -2{\rm e}^{-(r-R_0)/\alpha}\right],
\end{eqnarray}
where $R_0$ is the coordinate of the potential minimum, $\alpha$ is an effective range of interaction, $\gamma$ and $n_0$ are parameters of the model. Other two constants $C_{H}$ and $A$ are expressed via $\gamma$ and $n_0$ as follows
\begin{equation}
	C_{H} = \frac{n_0}{n_0 + \gamma - 2}, \quad A = \frac{2 - \gamma}{n_0},
\end{equation}
where $\varepsilon$ is the depth of the potential well at $r=R_0$. This potential is reduced to the ordinary Morse potential~\cite{Morse1929} at $\gamma=2$. For a more detailed discussion of such modified Morse potential, see Sections~1 and~2 in~\cite{KozlovskiiDobush2020}.

Modifications of the Morse potential have been used in other works as well. For example, in~\cite{MartinezValenciaEtAl2013} a repulsive term in a form of a power of $r^{-1}$ was added to the ordinary Morse potential, and the influence of the softness of such a term was investigated on the coordinates of the critical point. The generalized form of the Morse potential was suggested in~\cite{BiswasHamann1985}
\begin{equation}
	U(r) = A_1 {\rm e}^{-\lambda_1 r} + A_2 {\rm e}^{-\lambda_2 r}
\end{equation} 
with application to silicon structural energies, and was also considered in~\cite{Lim2005} as the potential for Be-S and H-Na compounds.

Our modification contains an additional repulsive term, similarly to~\cite{MartinezValenciaEtAl2013}, as well as introduces parameter $\gamma$, which can vary as opposed to being strictly equal to 2 in the Morse potential. Including the repulsive term enables us to single out a reference system (in the reciprocal space) and apply the method of collective variables to calculating the grand partition function~\cite{KozlovskiiDobush2016}.

\section{\label{sec:eos}The equation of state}
\subsection{Pressure as a function of temperature and chemical potential}
The equation of state obtained in~\cite{KozlovskiiDobush2020} reads
\begin{equation}
	\label{eq:eosMT}
	Pv\beta = E_\mu(M, T) + M \bar \rho_0 + \frac{1}{2} d \bar \rho_0^2 - \frac{a_4}{24} \bar \rho_0^4.
\end{equation}
The quantities in the left-hand side of the equation are $P$, the pressure; $\beta = (k_{\rm B} T)^{-1}$, the inverse temperature; $k_{\rm B}$, the Boltzmann constant; $T$, the temperature; $v$, cell volume. The quantities in the right-hand side are, in general, functions of the temperature $T$ and the chemical potential~$\mu$. Let us present their expressions explicitly.

First, the quantity $M$ depends linearly on the chemical potential
\begin{align}\label{chem_pot}
	&	M = \frac{\tilde\mu}{W(0)} + g_1 - \frac{g_3}{g_4} d - \frac{1}{6} \frac{g_3^3}{g_4^2}, \\
	&	\tilde\mu=\mu-\mu_0(1+\tau),
\end{align}
where $\mu_0$ is some positive constant, $\tau$ is the relative temperature $\tau = (T - T_c) / T_c$, $T_c$ is the critical temperature. We will call $M$ the effective chemical potential.

The quantity $W(0)$ is expressed via parameters of the potential~(\ref{def:mod_morse}) as follows
\begin{equation}
	W(0) = \Phi^{(r)}(0) \left[ B - 1 + \chi_0 + \tau (\chi_0 + A_\gamma) \right],
\end{equation}
where
\begin{align*} 
	& B = 2 \gamma^3 e^{(1-\gamma)R_0/\alpha},
	\nonumber \\
	& A_\gamma = A e^{(n_0-\gamma)R_0/\alpha} \left( \gamma / n_0\right)^3, 
\end{align*}
and $\Phi^{(r)}(0)$ is the Fourier transform of the repulsive part of the potential at $\abs{\vb k}=0$
\begin{equation*}
	\Phi^{(r)}(0) = \varepsilon C_H 8\pi {\rm e}^{\gamma R_0/\alpha} \left(\frac{\alpha}{\gamma R_0}\right)^3.
\end{equation*}

The parameter $\chi_0$ is used in~\cite{KozlovskiiDobush2020} to single out a contribution in the Fourier transform of the potential that is treated as a reference system defined in the reciprocal space, and is selected as $\chi_0 = 0.07$~\cite{KozlovskiiDobush2020}.

The coefficients $g_n$ are given by the formulas:
\begin{align}
	& g_0 = \ln T_0, \qquad g_1 = T_1/T_0, \qquad g_2 = T_2/T_0 - g_1^2,  
	\nonumber \\
	& g_3 = T_3/T_0 - g_1^3 - 3g_1 g_2, 
	\nonumber\\
	& g_4 = T_4/T_0 - g_1^4 - 6 g_1^2 g_2 - 4 g_1 g_3 - 3 g_2^2, 
	\nonumber\\
	& a_4 = -g_4,
\end{align}
where $T_n(p,\alpha^*)$ are the following special functions
\begin{equation}
	T_n(p,\alpha^*) = \sum_{m=0}^{\infty} \frac{(\alpha^*)^m}{m!} m^n {\rm e}^{-pm^2}.
\end{equation}
Here $\alpha^*=v e^{\beta_c\mu_0}$, and the parameter $p$ has the form
\begin{equation}
	p = \frac{\beta_c}{2} \Phi^{(r)}(0) [\chi_0 + A_\gamma].
\end{equation} 

The quantity $\beta_c$ denotes the critical value of the inverse temperature. In~\cite{KozlovskiiDobush2020} it was found that
\begin{equation*}
	\varepsilon\beta_c = 0.200, \quad \frac{k_{\rm B} T_c}{\varepsilon} = 4.995.
\end{equation*}
We also use the reduced temperature defined as $T^* = k_{\rm B}T/\varepsilon$, and thus its critical value $T^*_c = 4.995.$

Since $p$ is independent of temperature, the coefficients $g_n$ are also independent of temperature. The numerical values for other coefficients used in this paper are the same as those in~\cite[eqs.~(5), (23), and~(24)]{KozlovskiiDobush2020}:
\begin{eqnarray}
	\label{params}
	&& \chi_0 = 0.07,  \quad \gamma = 1.65, \nonumber\\
	&& n_0 = 1.521,  \quad R_0/\alpha = 2.9544, \nonumber\\
	&& \alpha^* = 5.0  \qquad p = 1.0.
\end{eqnarray}

The quantity $d$ entering equations~\eqref{eq:eosMT} and~\eqref{chem_pot} is a function of temperature
\begin{equation}
	\label{def:D0}
	d = g_2 - \frac{1}{2} \frac{g_3^2}{g_4} - \frac{1}{\beta W(0)}.
\end{equation}
The condition $d = 0$ defines the critical temperature~\cite{KozlovskiiDobush2020}
\begin{equation}
	k_{\rm B}T_c = \left(g_2 - \frac{1}{2} \frac{g_3^2}{g_4} \right) (B - 1 + \chi_0) \Phi^{(r)}(0).
\end{equation}  

The function $E_\mu(M, T)$ from the equation \eqref{eq:eosMT} is provided by
\begin{eqnarray}\label{eq:E_mu}
	E_\mu (M, T) & = & - \frac{\ln (2\pi \beta W(0))}{2 N_v}   +  g_0 - \frac{\beta W(0)}{2} 
	\left(\frac{\tilde\mu}{W(0)} \right)^{2} 
	\nonumber\\
	&& - \frac{g_3}{g_4} {M} \! - \frac{g_3^2}{2 g_4^2}  d - \frac{1}{24} \frac{g_3^4}{g_4^3}. 
\end{eqnarray}
Here the quantity $N_v$ defines the number of cubic cells in volume $V$ for the initial model.
In the thermodynamic limit, $N_v \to \infty$, and thus, the first term can be neglected. The term $\tilde{\mu}/W(0)$ can be expressed in terms of $M$ using~\eqref{chem_pot}. The temperature and the inverse temperature can always be expressed in terms of the reduced temperature and a corresponding critical value:
\begin{equation*}
	T = T_c(1+\tau), \quad \beta = \beta_c (1 + \tau)^{-1}.
\end{equation*} 

The quantity $\bar{\rho}_0$ is a solution to the following cubic equation
\begin{equation}\label{eq:ro_M}
	M + d \bar\rho_0 - \frac{a_4}{6} \bar\rho_0^3 = 0.
\end{equation}
For any $\tau > 0$, the latter equation has one real root
\begin{equation}\label{eq:ro_MT}
	\bar \rho_0 = \left(- \frac{3 M}{g_4} + \sqrt{Q_t}\right)^{1/3} - \left(  \frac{3 M}{g_4} + \sqrt{Q_t} \right)^{1/3},
\end{equation}
where
\begin{equation}
	Q_t = \left(  \frac{2d}{g_4}\right)^3 + \left( -\frac{3 M}{g_4}\right)^2, \qquad g_4<0.
\end{equation}
Thus, $\bar{\rho}_0$ is a function of the temperature and the chemical potential.

Let us introduce the reduced pressure
\begin{equation}
	\label{def:reduced_pres}
	P^* \equiv \frac{P v}{\varepsilon}.
\end{equation}
Considering the equation of state~\eqref{eq:eosMT}, $P^*$ is explicitly written as
\begin{eqnarray}
	\label{eq:eosPTM_reduced}
	P^* & = & (1 + \tau)T^*_c \bigg[ E_\mu(M, T)  + M \bar \rho_0 + \frac{1}{2} d \bar \rho_0^2 - \frac{a_4}{24} \bar \rho_0^4
	\bigg].
\end{eqnarray}
This equation can be easily represented graphically. Figure~\ref{fig1} illustrates the relationship between the reduced pressure $P^*$ and the effective chemical potential $M$ for various values of the relative temperature $\tau$, at and above the critical temperature. At the critical point $M=0, \tau=0$, $P^*_c = 1.606$.

\subsection{Pressure as a function of temperature and density}
\begin{figure}[h!]
	\centering
	\includegraphics[width=0.45\textwidth]{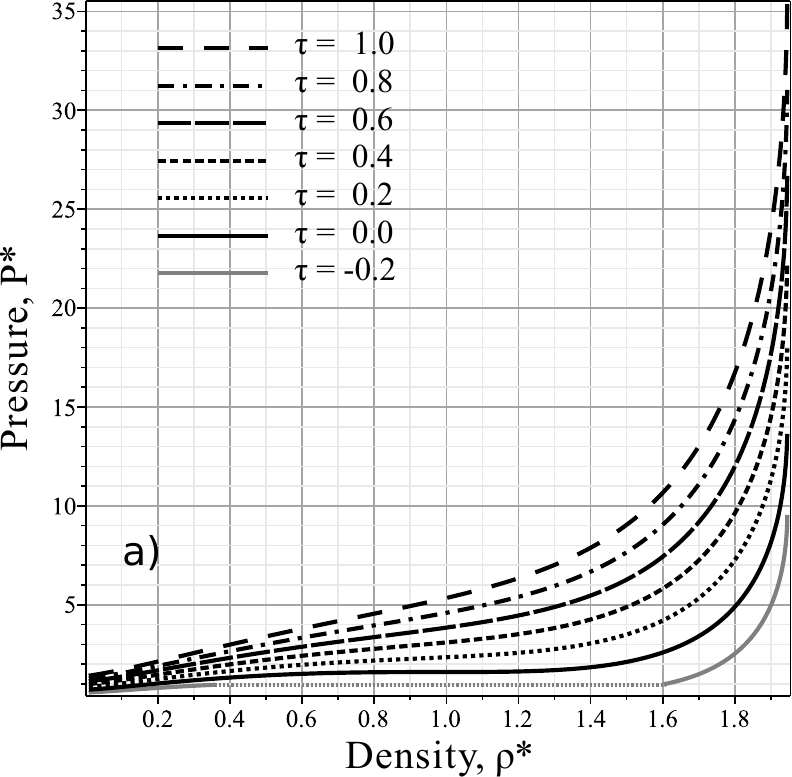} 
	\includegraphics[width=0.45\textwidth]{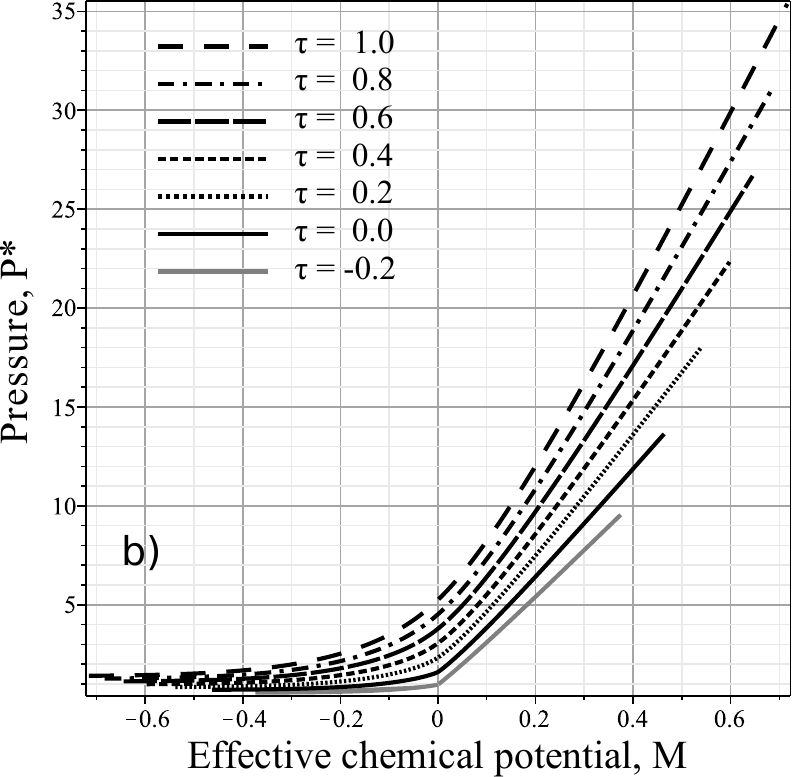} 
    \captionsetup{width=0.9\textwidth}
	\caption{Isotherms of the reduced pressure $P^*$ as a function (a) of the density $\rho^*$, and (b) of the effective chemical potential $M$ at $T \geq T_c$ represented by black lines. Thick grey lines on both figures correspond to isotherms of pressure at $T < T_c$ based on the results taken from~\cite{KozlovskiiDobush2020}.
	}
	\label{fig1}
\end{figure}

In this work, by density we mean the particle number density $\rho = \langle N \rangle / V$.
In the framework of the grand canonical ensemble the average number of particles $\left\langle N \right\rangle$ is found by 
\begin{eqnarray*}
	\langle N \rangle & = & \left(\frac{\partial \ln\Xi}{\partial \beta \mu}\right)_{T,V} 
	\\
	& = & V \left(\frac{\partial P}{\partial \mu}\right)_{T,V}.
\end{eqnarray*}
From this equation it follows
\begin{equation*}
	\frac{\langle N \rangle}{V} = \frac{\langle N \rangle}{v N_v} = \left(\frac{\partial P}{\partial \mu}\right)_{T,V}
\end{equation*}
and thus
\begin{equation}
	\rho^* \equiv \frac{\langle N \rangle}{V} v = v \left(\frac{\partial P}{\partial \mu}\right)_{T,V} 
	= \left(\frac{\partial (Pv\beta)}{\partial (\beta \mu)}\right)_{T,V}.
\end{equation}
The quantity $\rho^*$, on the one hand, is the reduced particle number density, which is the notation commonly used in the literature on simple fluids~\cite{HansenMcDonald2013}. In the context of the cell model, on the other hand, it is the average number of particles per cell.\footnote{In our previous works we denoted the reduced number density by $\bar{n}$. In the current work we switch to more common notation $\rho^*$.}
Taking explicit derivatives, we arrive at
\begin{equation}\label{eq:density}
	\rho^* = \rho^*_c - M + \frac{ \bar \rho_0}{\beta W(0)}.
\end{equation}
The quantity $\rho^*_c$ in the equation \eqref{eq:density} is the critical density~\cite{KozlovskiiDobush2020}
\begin{eqnarray}\label{eq:crit_dens}
	\rho^*_c & = & g_1 - \frac{g_3}{g_4}\left(  g_2 - \frac{1}{2} \frac{g_3^2}{g_4}\right) - \frac{1}{6} \frac{g_3^3}{g_4^2} 
	\nonumber\\
	& = & g_1 - \frac{g_2 g_3}{g_4} + \frac{g_3^3}{3g_4^2}
\end{eqnarray}
Its numerical value for parameters~\eqref{params} is
\begin{equation*}
	\rho^*_c = 0.978.
\end{equation*}
Equations~\eqref{eq:density} and~\eqref{eq:eosMT} jointly define a parametric relationship between pressure and density, with $M$ serving as the parameter. Figure~\ref{fig1} shows the dependence of the reduced pressure $P^*$ on the density $\rho^*$ for various values of the reduced temperature $\tau$, at and above the critical temperature.

This dependence can also be expressed explicitly. To achieve this, we combine the equations~\eqref{eq:density} and~\eqref{eq:ro_M} to express the effective chemical potential $M$ as a function of density and temperature:
\begin{equation}\label{eq:M_nT}
	\bar M = \frac{\rho_{n}}{\beta W(0)} - (\rho^* - \rho^*_c),
\end{equation}
where
\begin{align} \label{eq:ro_nT}
	& \rho_{n} = - 2 \left(\frac{g_3^2 - 2g_2 g_4}{g_4^2} \right)^{\frac{1}{2}} \cos \left( \frac{\alpha_n}{3} + \frac{\pi}{3} \right), \\
	& \alpha_n = \arccos \left[ \left( - \frac{9 g_4^4}{\left( 2 g_2 g_4 - g_3^2\right)^3}\right)^{\frac{1}{2}} (\rho^*_c - \rho^*)\right]. \nonumber 
\end{align}
The notation $\bar{M}$ represents the effective chemical potential as a function of temperature and density $\rho^*$, while $M$ represents the effective chemical potential as a function of temperature and chemical potential~$\mu$.

At $T>T_c$ the equation of state of a cell fluid model in terms of the density and the temperature has the following form
\begin{equation}\label{eq:eosNT}
	Pv\beta = E_\rho (\rho^*,T) + \bar M \rho_{n} + \frac{d}{2} \rho_{n}^2 - \frac{a_4}{24} \rho_{n}^4.
\end{equation}
The quantity $E_\rho (\rho^*, T)$ in \eqref{eq:eosNT} is the same as function $E_\mu (M,T)$ \eqref{eq:E_mu} rewritten in terms of density and temperature taking into account the expression \eqref{eq:M_nT}
\begin{eqnarray}\label{eq:E_nu}
	E_\rho (\rho^*, T) & = & - \frac{\ln (2\pi \beta W(0))}{2 N_v}  +  g_0  - \frac{\beta W(0)}{2} 
	\left(\bar{M} - g_1 + \frac{g_3}{g_4} d + \frac{g_3^3}{6g_4^2} \right)^{2} 
	\nonumber\\
	&& - \frac{g_3}{g_4} {\bar{M}} - \frac{g_3^2}{2 g_4^2}  d - \frac{1}{24} \frac{g_3^4}{g_4^3}. 
\end{eqnarray}
For the reduced pressure we write
\begin{eqnarray}
	\label{eq:eosPTn_reduced}
	P^* & = & (1 + \tau)T^*_c \bigg[ E_\rho (\rho^*,T)  + \bar{M} \rho_{n} + \frac{d}{2} \rho_{n}^2 - \frac{a_4}{24} \rho_{n}^4\bigg].
\end{eqnarray}

The equations of state~\eqref{eq:eosMT} and \eqref{eq:eosNT} are derived in the zero-mode approximation of the $\rho^4$-model, which imposes limits on their applicability. Specifically, in terms of density, the equations are  applicable for $\rho^*_{\rm min} \leq \rho^* < \rho^*_{\rm max}$, where $\rho^*_{\rm min}$ and $\rho^*_{\rm max}$ are determined by the parameters $\alpha^*$ and $p$. For parameters given in~\eqref{params}, these values were estimated in~\cite{KozlovskiiDobush2020} as $\rho^*_{\rm min} = 0.009$ and $\rho^*_{\rm max} = 1.946$.

Figure~\ref{fig1} shows the isotherms for the pressure $P^*$ as a function of density $\rho^*$ (see Fig.~\ref{fig1}a), and the effective chemical potential $M$ (see Fig.~\ref{fig1}b).

Thus, in this Section, we presented two forms of the equation of state. The first one expresses the pressure as a function of temperature and chemical potential, $P = P(T, \mu)$, and is represented by equivalent Eqs.~\eqref{eq:eosMT} and~\eqref{eq:eosPTM_reduced}. The second one expresses the pressure as a function of temperature and density, $P = P(T, \rho^*)$, and is represented by equivalent Eqs.~\eqref{eq:eosNT} and~\eqref{eq:eosPTn_reduced}. These equations, together with the explicit dependence of density on temperature and chemical potential, $\rho^* = \rho^*(T, \mu)$, Eq.~\eqref{eq:density}, are the basis for the calculation of the thermodynamic response functions in the next Section~\ref{sec:respons_functions}. 

\section{\label{sec:respons_functions}Thermodynamic response functions}

\subsection{Isothermal compressibility}
The isothermal compressibility is defined by
\begin{equation}
	\label{def:isotherm_compres}
	\kappa_T = -\frac{1}{V}\left(\frac{\partial V}{\partial P}\right)_{T, N}.
\end{equation}
Let us perform some transformations to rewrite $\kappa_T$ into a form that is more suitable for the equation of state $P=P(T,\mu)$, see~\eqref{eq:eosMT} and~\eqref{eq:eosPTM_reduced}:
\begin{eqnarray*}
	\kappa_T	& = & \frac{1}{\rho} \left(\frac{\partial \rho}{\partial P} \right)_{T,N}
	\\
	& = & \frac{1}{\rho} \frac{\left(\partial \rho / \partial \mu\right)_{T}}
	{\left(\partial P / \partial \mu\right)_{T}}.
\end{eqnarray*}
We have omitted the condition of constant $N$ in the last line of the above equation since we have explicit dependencies on temperature and chemical potential for both pressure $P = P(T, \mu)$, Eq.~\eqref{eq:eosMT}, and density $\rho = \rho(T, \mu)$, Eq.~\eqref{eq:density}.
Applying the Gibbs--Duhem equation
\begin{equation}
	N{\rm d}\mu = -S{\rm d}T + V{\rm d}P,
\end{equation}
at $T = const$ one has 
\begin{equation*}
	{\rm d} P = \rho {\rm d} \mu,
\end{equation*}
or
\begin{equation}
	\label{eq:rho_dpdm}
	\rho = \left(\frac{\partial P}{\partial \mu}\right)_T.
\end{equation}
Substituting this into the last expression for $\kappa_T$, one gets
\begin{equation}
	\kappa_T = \frac{1}{\rho^2} \left(\frac{\partial \rho}{\partial \mu}\right)_{T} .
\end{equation}
Finally, from~\eqref{eq:rho_dpdm} it follows that
\begin{equation}
	\left(\frac{\partial \rho}{\partial \mu}\right)_T = \left(\frac{\partial^2 P}{\partial \mu^2}\right)_T ,
\end{equation}
and ultimately we arrive at the very useful expression for the isothermal compressibility
\begin{equation}
	\kappa_T = \frac{1}{\rho^2} \left(\frac{\partial^2 P}{\partial \mu^2}\right)_{T} .
\end{equation}

Let us introduce the reduced isothermal compressibility
\begin{equation}
	\kappa^*_T \equiv \frac{\varepsilon K_T}{v}.
\end{equation}
The quantity $\kappa^*_T$ is dimensionless and is of order unity, except at the critical point itself, where it is divergent. It is expressed in terms of the reduced quantities $P^*$ and $\rho^*$ as follows
\begin{eqnarray}
	\label{eq:kappa_star_m1}
	\kappa^*_T & = & \frac{\varepsilon}{{\rho^*}^2} \left(\frac{\partial \rho^*}{\partial \mu}\right)_T
	\\
	\label{eq:kappa_star_m}
	& = & \frac{\varepsilon^2}{{\rho^*}^2} \left(\frac{\partial^2 P^*}{\partial \mu^2}\right)_T.
\end{eqnarray}
Now, either Eq.~\eqref{eq:eosPTM_reduced} or Eq.~\eqref{eq:density} can be used to explicitly calculate $\kappa^*_T$, with the result expressed as a function of temperature and chemical potential. In Appendix~\ref{sec:app-a}, we provide the derivation of the explicit expression for the isothermal compressibility based on~\eqref{eq:kappa_star_m1}.

If it is preferable to use the equation of state in the form $P=P(T, \rho^*)$, see~\eqref{eq:eosNT} and~\eqref{eq:eosPTn_reduced}, then the most suitable expressions for the isothermal compressibility are
\begin{equation}
	\kappa_T = \frac{1}{\rho^*} \left(\frac{\partial P}{\partial \rho^*}\right)^{-1}_T
\end{equation}
and 
\begin{equation}
	\kappa^*_T = \frac{1}{\rho^*} \left(\frac{\partial P^*}{\partial \rho^*}\right)^{-1}_T.
\end{equation}

\begin{figure}[htbp]
	\centering
	\includegraphics[width=0.465\textwidth]{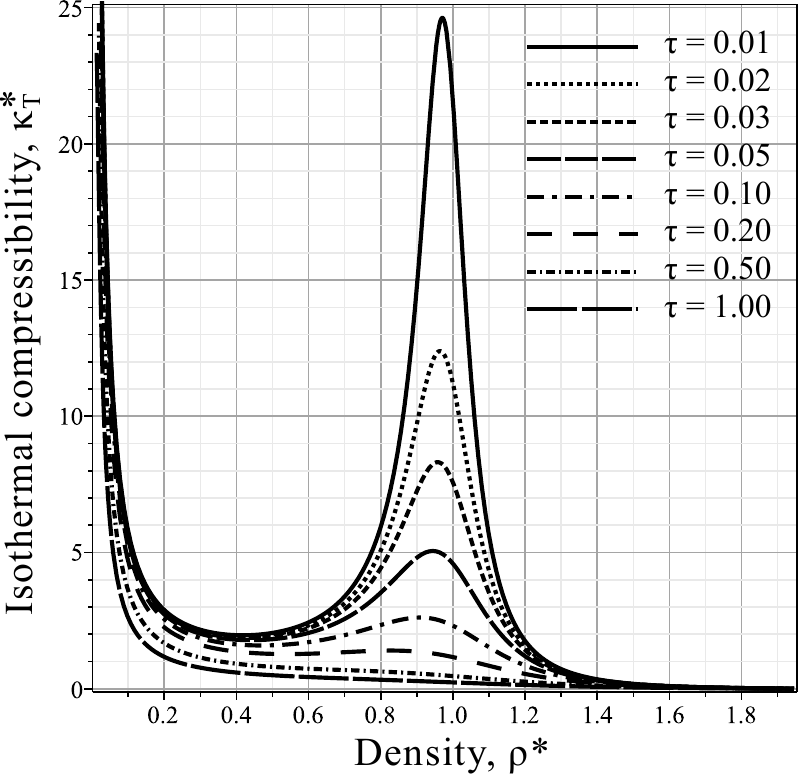}
    \captionsetup{width=0.9\textwidth}
	\caption{The reduced isothermal compressibility $\kappa^*_T$ as a function of the density $\rho^*$ at different values of reduced temperature $\tau > 0$ ($T > T_c$). 
	}
	\label{fig2a}
\end{figure}

\begin{figure}[h!] 
	\centering
	\includegraphics[width=0.45\textwidth]{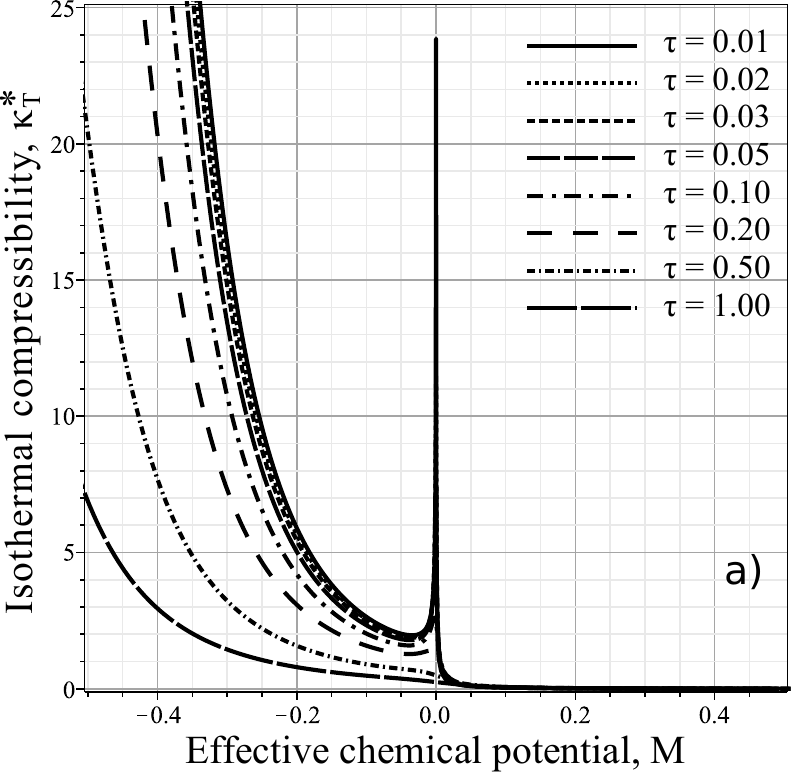}
	\includegraphics[width=0.461\textwidth]{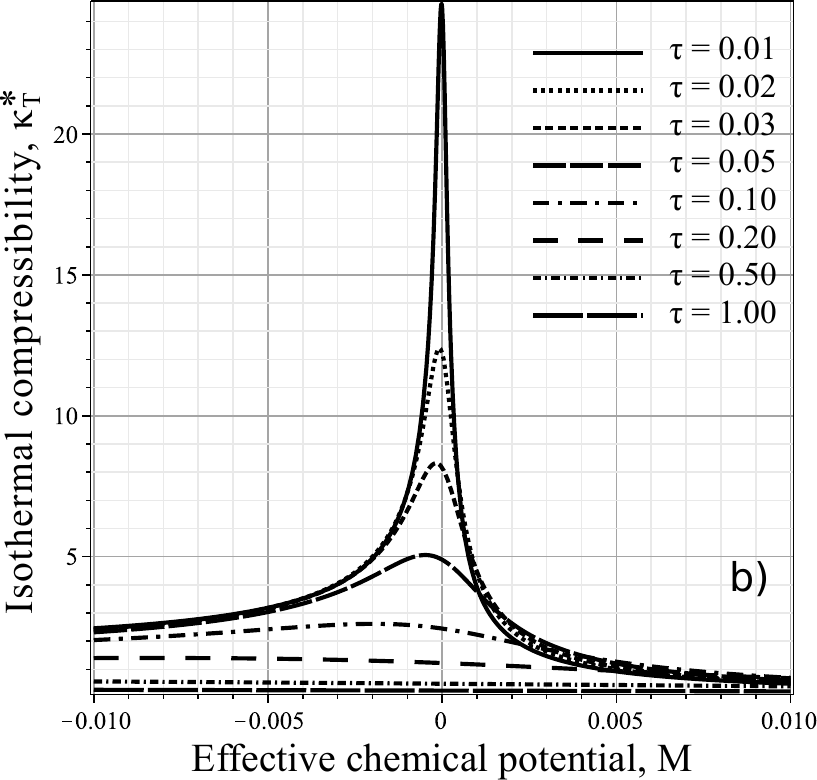}
	\captionsetup{width=0.9\textwidth}
	\caption{The reduced isothermal compressibility $\kappa^*_T$ as a function of the effective chemical potential $M$ for different temperatures $\tau = (T - T_c)/T_c$ at $T > T_c$. The two figures differ in the scale of $M$. Part~(a) covers a wider range of $M$. Part~(b) focuses on a range of $M$ around its critical value $0$.
	}
	\label{fig2b}
\end{figure}

Figure~\ref{fig2a} illustrates the dependence of the compressibility $\kappa^*_T$ on the density $\rho^*$ for various values of temperature above the critical one. The dependence of $\kappa^*_T$ on the effective chemical potential $M$ is displayed in Fig.~\ref{fig2b}.

\subsection{Thermal pressure coefficient}
The thermal pressure coefficient is defined by
\begin{equation}
	\label{def:therm_pres_coef}
	\beta_V = \left( \frac{\partial P}{\partial T} \right)_{V,N}.
\end{equation}
We rewrite $\beta_V$ into a form that is suitable for the equation of state $P=P(T,\mu)$, see~\eqref{eq:eosMT} and~\eqref{eq:eosPTM_reduced},
\begin{eqnarray}
	\beta_V	& = & \left(\frac{\partial P}{\partial T}\right)_{\mu} + \left(\frac{\partial P}{\partial \mu}\right)_T \left(\frac{\partial \mu}{\partial T}\right)_{V, N}.
\end{eqnarray}
Applying the cyclic relation
\begin{equation*}
	\left(\frac{\partial \mu}{\partial T}\right)_{V, N} 
	\left(\frac{\partial T}{\partial V}\right)_{\mu, N}
	\left(\frac{\partial V}{\partial \mu}\right)_{T, N}
	= -1
\end{equation*}
we obtain
\begin{eqnarray*}
	\left(\frac{\partial \mu}{\partial T}\right)_{V, N} 
	& = & 
	- \left(\frac{\partial V}{\partial T}\right)_{\mu, N} 
	\left(\frac{\partial V}{\partial \mu}\right)^{-1}_{T, N} \\
	& = & - \left(\frac{\partial \rho^*}{\partial T}\right)_{\mu}
	\left(\frac{\partial \rho^*}{\partial \mu}\right)^{-1}_{T}.		
\end{eqnarray*}
Substituting this result into the formula for $\beta_V$, we arrive at the final expression for the thermal pressure coefficient
\begin{equation}
	\beta_V = \left(\frac{\partial P}{\partial T}\right)_{\mu} 
	- \left(\frac{\partial P}{\partial \mu}\right)_T 
	\left(\frac{\partial \rho^*}{\partial T}\right)_{\mu}
	\left(\frac{\partial \rho^*}{\partial \mu}\right)^{-1}_{T}
\end{equation}
which\footnote{Compare this equation for $\beta_V$ with Eq.(17) from~\cite{StrokerMeier2021}} is easily calculated based on Eqs.~\eqref{eq:eosMT} and~\eqref{eq:density}. It is also worth noting that the first contribution to $\beta_V$ is essentially the entropy per volume, $S/V = (\partial P / \partial T)_\mu$.

We introduce the reduced thermal pressure coefficient by
\begin{equation}
	\beta^*_V = \frac{v}{k_{\rm B}}\beta_V.
\end{equation}
It is expressed in terms of reduced quantities as follows
\begin{equation}
	\label{eq:beta_star_m}
	\beta^*_V = \frac{1}{T^*_c} 
	\left[ \left(\frac{\partial P^*}{\partial \tau}\right)_{\mu} 
	- \left(\frac{\partial P^*}{\partial \mu}\right)_T 
	\left(\frac{\partial \rho^*}{\partial \tau}\right)_{\mu}
	\left(\frac{\partial \rho^*}{\partial \mu}\right)^{-1}_{T} 
	\right].
\end{equation}

If it is preferable to use the equation of state in the form $P=P(T, \rho^*)$, see~\eqref{eq:eosNT} and~\eqref{eq:eosPTn_reduced}, then the most suitable expressions for the thermal pressure coefficient follow immediately from the definition~\eqref{def:therm_pres_coef}
\begin{equation}
	\beta_V = \left(\frac{\partial P}{\partial T}\right)_\rho
\end{equation}
and
\begin{equation}
	\label{eq:beta_star_n}
	\beta^*_V = \frac{1}{T^*_c} \left(\frac{\partial P^*}{\partial \tau}\right)_{\rho}.
\end{equation}
Explicit calculation for $\beta^*_V$ is presented in Appendix~\ref{sec:app-a}.

\begin{figure}[h!] 
	\centering
	\includegraphics[width=0.45\textwidth]{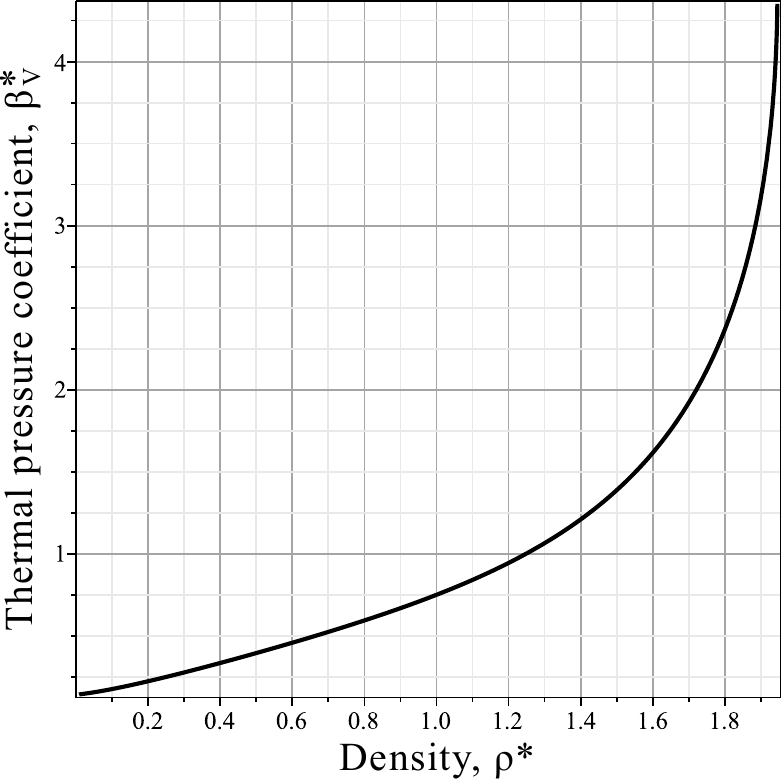}
	\captionsetup{width=0.9\textwidth}
	\caption{The reduced thermal pressure coefficient $\beta^*_V$ as a function of the density $\rho^*$ at different values of relative temperature $\tau > 0$~($T > T_c$). Multiple isotherms collapsing onto the same line, making them indistinguishable at the scale of the figure.
	}
	\label{fig3a}
\end{figure}
\begin{figure}[h!]
	\centering
	\includegraphics[width=0.45\textwidth]{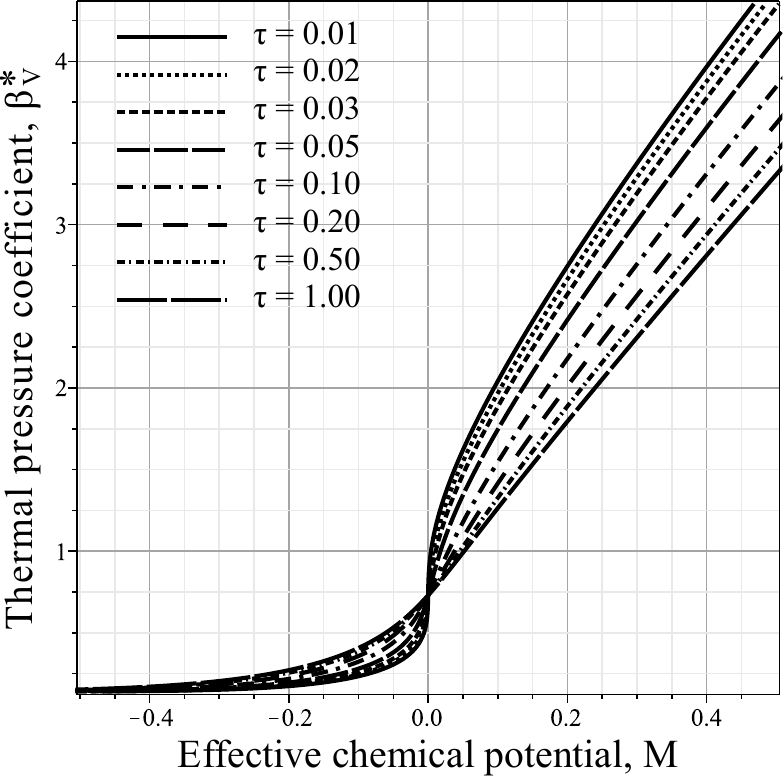}
	\captionsetup{width=0.9\textwidth}
	\caption{The reduced thermal pressure coefficient $\beta^*_V$ as a function of the effective chemical potential $M$ at different values of relative temperature $\tau > 0$~($T > T_c$). 
	}
	\label{fig3b}
\end{figure}
\begin{figure}[h!]
	\centering
	\includegraphics[width=0.45\textwidth]{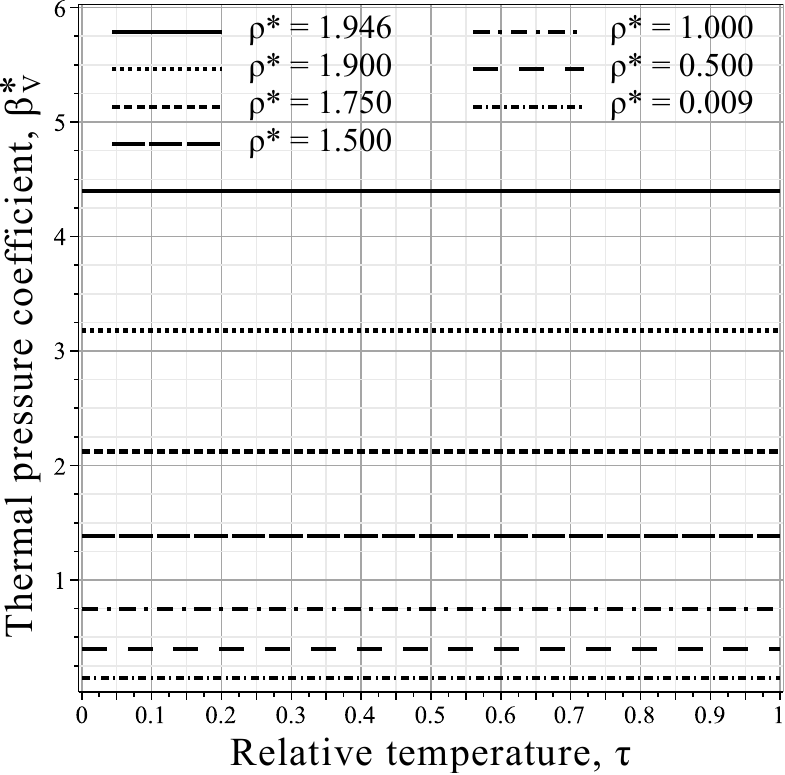}
	\captionsetup{width=0.9\textwidth}
	\caption{The reduced thermal pressure coefficient $\beta^*_V$ as a function of the relative temperature $\tau$ at different values of the density $\rho^*$. 
	}
	\label{fig3c}
\end{figure}

Figure~\ref{fig3a} shows the dependence of the pressure coefficient $\beta^*_V$ on the density $\rho^*$ for various values of temperature above the critical one. It is interesting to note that the temperature dependence is very weak in this case, and multiple isotherms collapse onto the same line and are indistinguishable at the scale of the figure. This is not a surprised behavior, as similar one is observed for the thermal pressure coefficient of the Lennard-Jones fluid~\cite{YigzaweSadus2013} as well. The dependence of $\beta^*_V$ on the effective chemical potential $M$ is displayed in Fig.~\ref{fig3b}. The dependence of $\beta^*_V$ on the relative temperature $\tau$ is displayed in Fig.~\ref{fig3c} for a few values of density.

\subsection{Thermal expansion coefficient}
The thermal expansion coefficient is defined by
\begin{equation}
	\alpha_P = \frac{1}{V}\left(\frac{\partial V}{\partial T}\right)_{P,N}.
\end{equation}
We rewrite $\alpha_P$ into a form suitable for the equation of state $P=P(T,\mu)$, see~\eqref{eq:eosMT} and~\eqref{eq:eosPTM_reduced}
\begin{eqnarray}
	\label{eq:alpha_vs_rho}
	\alpha_P & = & -\frac{1}{\rho} \left(\frac{\partial \rho}{\partial T}\right)_{P, N}
	\\
	& = & -\frac{1}{\rho} \left(\frac{\partial \rho}{\partial T}\right)_{\mu} 
	- \frac{1}{\rho} \left(\frac{\partial \rho}{\partial \mu}\right)_{T}
	\left(\frac{\partial \mu}{\partial T}\right)_{P, N}.
\end{eqnarray}
Applying the cyclic relation
\begin{equation*}
	\left(\frac{\partial \mu}{\partial T}\right)_{P, N}
	\left(\frac{\partial T}{\partial P}\right)_{\mu, N}
	\left(\frac{\partial P}{\partial \mu}\right)_{T, N}
	= -1
\end{equation*}
we get
\begin{eqnarray*}
	\left(\frac{\partial \mu}{\partial T}\right)_{P, N} & = & 
	- \left(\frac{\partial P}{\partial T}\right)_{\mu}
	\left(\frac{\partial P}{\partial \mu}\right)^{-1}_{T}.
\end{eqnarray*}
Substituting the result into the expression for $\alpha_P$, we arrive at the final formula for the thermal expansion coefficient
\begin{equation}
	\label{eq:alphaP1}
	\alpha_P = -\frac{1}{\rho}\left(\frac{\partial \rho}{\partial T}\right)_{\mu}
	+ \frac{1}{\rho} \left(\frac{\partial \rho}{\partial \mu}\right)_{T}
	\left(\frac{\partial P}{\partial T}\right)_{\mu}
	\left(\frac{\partial P}{\partial \mu}\right)^{-1}_{T}
\end{equation}
which\footnote{Compare this equation for $\alpha_P$ with Eqs.(57)-(58) from~\cite{StrokerMeier2021}} is readily calculated based on Eqs.~\eqref{eq:eosMT} and~\eqref{eq:density}.

We introduce the reduced thermal expansion coefficient by
\begin{equation}
	\alpha^*_P = \frac{\varepsilon}{k_{\rm B}} \alpha_P.
\end{equation}
It is expressed in terms of reduced quantities as follows
\begin{eqnarray}
	\alpha^*_P & = & \frac{1}{T^*_c \rho^*}
	\left[ 
	-\left(\frac{\partial \rho^*}{\partial \tau}\right)_{\mu}
	\right. \nonumber\\
	&& + \left.\left(\frac{\partial \rho^*}{\partial \mu}\right)_{T}
	\left(\frac{\partial P^*}{\partial \tau}\right)_{\mu}
	\left(\frac{\partial P^*}{\partial \mu}\right)^{-1}_{T} 
	\right].
\end{eqnarray}

If it is preferable to use the equation of state in the form $P=P(T, \rho^*)$, see~\eqref{eq:eosNT} and~\eqref{eq:eosPTn_reduced}, then we need to rewrite the definition for $\alpha_P$ accordingly. We start with~\eqref{eq:alpha_vs_rho} and apply the cyclic relation
\begin{equation}
	\left(\frac{\partial \rho}{\partial T}\right)_P = - \left(\frac{\partial P}{\partial T}\right)_{\rho} \left(\frac{\partial P}{\partial \rho}\right)^{-1}_T
\end{equation}
to get for $\alpha_P$
\begin{equation}
	\alpha_P = \frac{1}{\rho} \left(\frac{\partial P}{\partial T}\right)_{\rho} \left(\frac{\partial P}{\partial \rho}\right)^{-1}_T ,
\end{equation}
or for $\alpha^*_P$
\begin{equation}
	\alpha^*_P = \frac{1}{\rho^* T^*_c} \left(\frac{\partial P^*}{\partial \tau}\right)_{\rho} \left(\frac{\partial P^*}{\partial \rho^*}\right)^{-1}_T.
\end{equation}

\begin{figure}[h!] 
	\centering
	\includegraphics[width=0.45\textwidth]{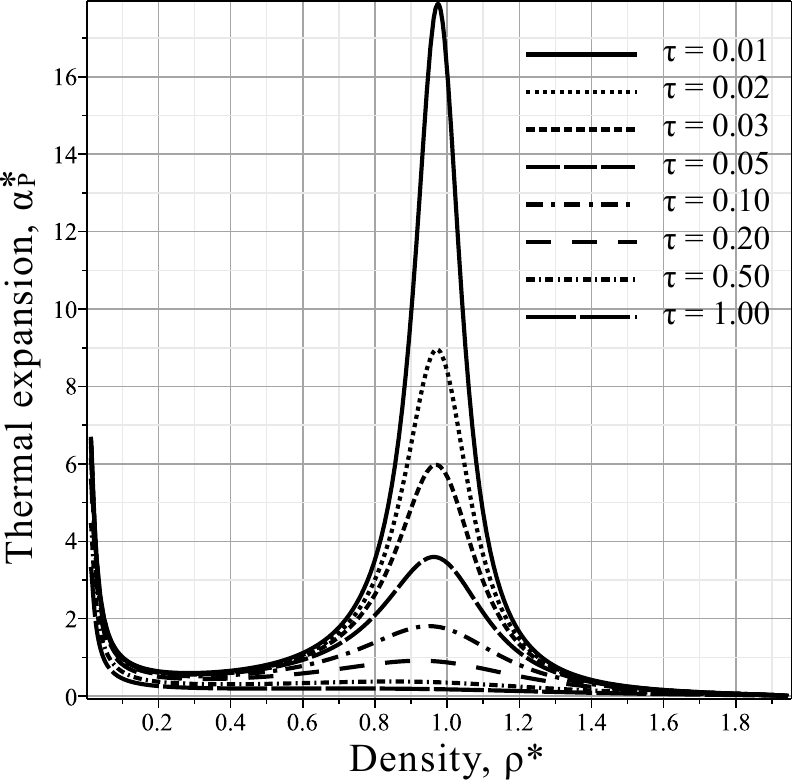}
	\captionsetup{width=0.9\textwidth}
	\caption{The reduced thermal expansion coefficient $\alpha^*_P$ as a function of the density $\rho^*$ at different values of relative temperature $\tau > 0$~($T > T_c$). 
	}
	\label{fig4a}
\end{figure}

\begin{figure}[h!]
	\centering
	\includegraphics[width=0.45\textwidth]{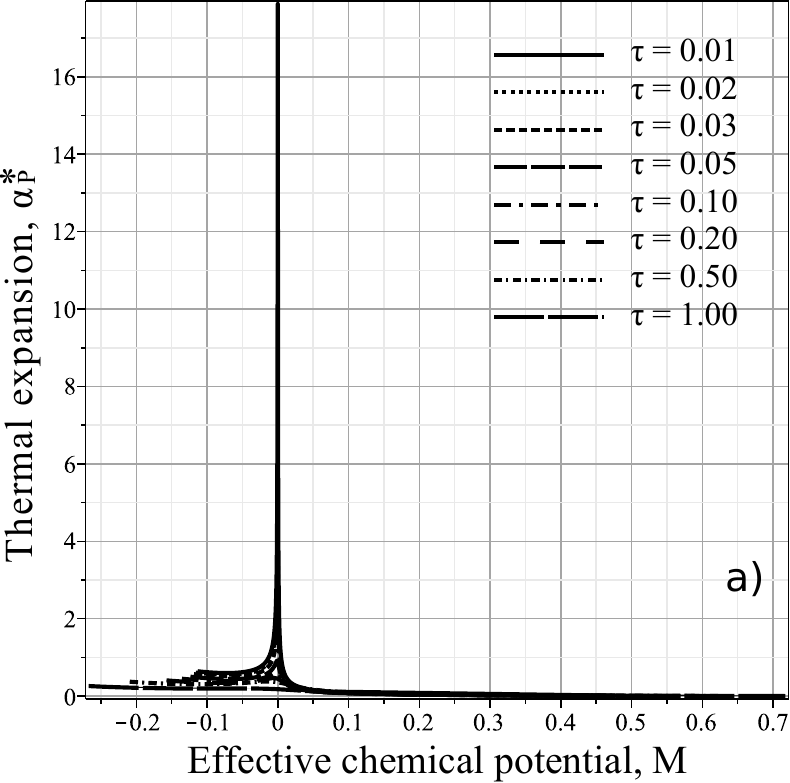}
	\includegraphics[width=0.471\textwidth]{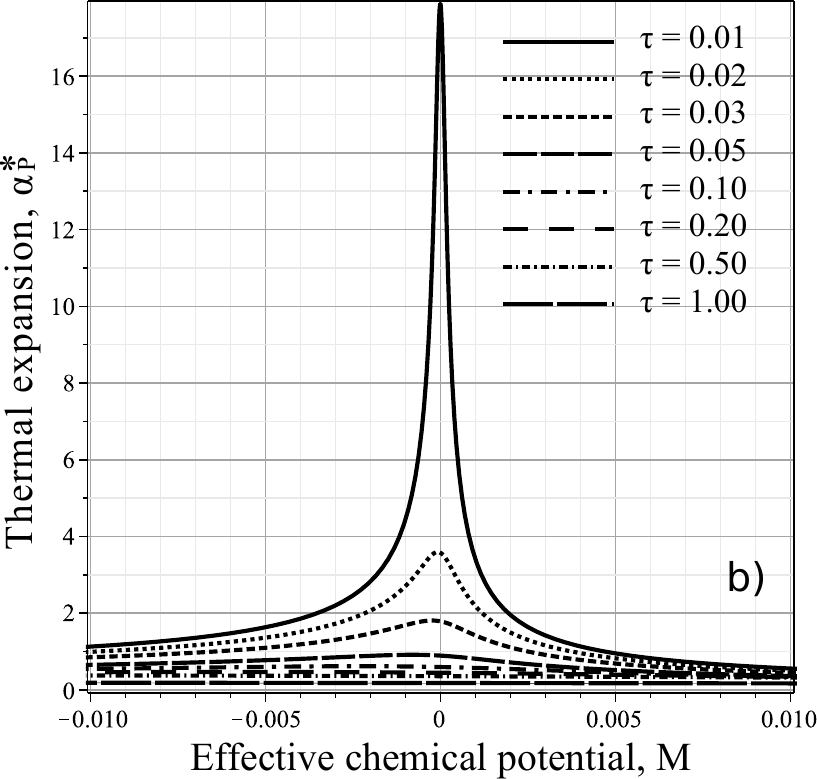}
	\captionsetup{width=0.9\textwidth}
	\caption{The reduced thermal expansion coefficient $\alpha^*_P$ as a function of the effective chemical potential $M$ for different temperatures $\tau = (T - T_c)/T_c$ at $T > T_c$. The two figures differ in the scale of $M$. Part (a) covers a wider range of $M$. Part (b) focuses on a range of $M$ around its critical value~$0$.
	}
	\label{fig4b}
\end{figure}

Figure~\ref{fig4a} shows the dependence of the reduced thermal expansion coefficient $\alpha^*_P$ on the density $\rho^*$ for various values of temperature above the critical one. The dependence of $\alpha^*_P$ on the effective chemical potential $M$ is displayed in Fig.~\ref{fig4b}.

\subsection{Relation between response functions}
The following thermodynamic identity holds between the calculated response functions
\begin{equation}
	\label{eq:identity}
	\frac{\alpha_P}{\kappa_T \beta_V} = \frac{\alpha^*_P}{\kappa^*_T \beta^*_V} = 1,
\end{equation}
which is easily derived from the cyclic relation between $P$, $V$ and $T$. In our case, this equality is exactly reproduced by substituting explicit expressions for $\kappa_T$, $\alpha_P$, and $\beta_V$ obtained from the equation of state~\eqref{eq:eosMT} (or~\eqref{eq:eosNT}). 


\section{Conclusions}
Thermodynamic response functions, namely the isothermal compressibility, the thermal pressure coefficient, and the thermal expansion coefficient, are calculated for a many-particle system interacting through a modified Morse potential. The starting point for these calculations are the equation of state obtained for a cell fluid model within the framework of the grand canonical ensemble in our previous work~\cite{KozlovskiiDobush2020}. The dependencies of the calculated response functions on the density and the effective chemical potential are illustrated graphically. The thermodynamic identity~\eqref{eq:identity} among these quantities is verified to confirm the self-consistency of the performed calculations.

\vskip3mm \textit{This work was supported by the National Research Foundation of Ukraine under the project No.~2023.03/0201.}

\appendix
\renewcommand{\theequation}{A.\arabic{equation}}
\setcounter{equation}{0}

\section{\label{sec:app-a} Explicit expressions for derivatives}
This Appendix contains two examples of explicit derivation of response functions. 

First, we provide an explicit example of differentiating the equation of state, namely~\eqref{eq:eosPTn_reduced}, by deriving the expression for the thermal pressure coefficient $\beta^*_V$ from~\eqref{eq:beta_star_n}.
For the quantity $\bar{M}$ from Eq.~\eqref{eq:M_nT} one has
\begin{equation}
	\left( \frac{\partial \bar{M}}{\partial \tau} \right)_{\rho^*} =
	\frac{k_{\rm B}T_c \rho_n}{W(0)} \frac{1 - \omega_0}{1 + \tau\omega_0},
\end{equation}
where
\begin{equation*}
	\omega_0 = \frac{\chi_0 + A_{\gamma}}{B - 1 + \chi_0}.
\end{equation*}
For $d$ from~\eqref{def:D0} one has
\begin{equation}
	\frac{\partial d}{\partial \tau} = -\frac{k_{\rm B}T_c}{W(0)} \frac{1 - \omega_0}{1 + \tau \omega_0}.
\end{equation}
While calculating both derivatives, we take into account
\begin{equation}
	\frac{\partial}{\partial \tau} \left[\beta W(0)\right] = -\frac{\beta_c \Phi^{(r)}(0)}{(1+\tau)^2} (B - 1 - A_{\gamma}).
\end{equation}
The derivative of the quantity $E_\rho$ from~\eqref{eq:E_nu} with respect to temperature $\tau$ is
\begin{eqnarray}
	\frac{\partial E_\rho}{\partial \tau} & = &  -\frac{1}{2}\left(\bar{M} - g_1 + \frac{g_3}{g_4} d + \frac{g_3^3}{6g_4^2} \right)^{2} 
	\frac{\partial \beta W(0)}{\partial \tau} -\beta W(0)\left(\bar{M} - g_1 + \frac{g_3}{g_4} d + \frac{g_3^3}{6g_4^2} \right)
	\nonumber\\
	&& \times \left(\frac{\partial \bar{M}}{\partial \tau} + \frac{g_3}{g_4}\frac{\partial d}{\partial \tau}\right) -\frac{g_3}{g_4}\frac{\partial \bar{M}}{\partial \tau} 
	- \frac{g_3^2}{2g_4^2}\frac{\partial d}{\partial \tau}.
\end{eqnarray}
Collecting these all formulas together, we can explicitly calculate $\beta^*_V$ from~\eqref{eq:beta_star_n}
\begin{eqnarray}
	\beta^*_V & = & E_\rho (\rho^*,T) + \bar{M} \rho_{n} + \frac{d}{2} \rho_{n}^2 - \frac{a_4}{24} \rho_{n}^4 + (1+\tau) \left(\frac{\partial E_{\rho}}{\partial \tau} + \frac{\partial \bar{M}}{\partial \tau}\rho_{n} + \frac{\rho_n^2}{2}\frac{\partial d}{\partial \tau}\right).
\end{eqnarray}
Substituting the expressions for derivatives and grouping similar terms, one finally arrives at
\begin{eqnarray}
	\beta^*_V & = & E_\rho (\rho^*,T) + \bar{M} \rho_{n} + \frac{d}{2} \rho_{n}^2 - \frac{a_4}{24} \rho_{n}^4 +  \left\{ \frac{\beta}{2}  \Phi^{(r)}(0) [B - 1 - A_\gamma]	G_M^2 \right . \nonumber \\
	& + & \left. \frac{1}{\beta W(0)} \frac{1 - \omega_0}{1 + \tau \omega_0} \left[ \frac{g_3^2}{2 g_4^2} + \frac{\rho_n^2}{2} - \frac{g_3}{g_4} \rho_n  +  \beta W(0) G_M \left( \frac{g_3}{g_4} - \rho_n \right)    \right] \right\},
\end{eqnarray}
where we introduced notation
\begin{equation*}
	G_M = \bar{M} - g_1+ \frac{g_3}{g_4} d + \frac{1}{6} \frac{g_3^3}{g_4^2}.
\end{equation*}

Second, we derive the explicit expression for the isothermal compressibility $\kappa^*_T$ using~\eqref{eq:kappa_star_m}, taking the density $\rho^*$ from \eqref{eq:density}.
For the effective chemical potential $M$ from \eqref{chem_pot} one has
\begin{equation}
	\left(\frac{\partial M}{\partial \mu} \right)_T = \frac{1}{W(0)}.
\end{equation}
For the quantity $\bar \rho_0$ from \eqref{eq:ro_MT} one has
\begin{equation}
	\left(\frac{\partial \bar \rho_0}{\partial \mu} \right)_T = \frac{1}{ g_4 W(0) \sqrt{Q_t}} \left[ \bar \rho_0 - 2 \left( - \frac{3M}{g_4} + \sqrt{Q_t} \right)^{\frac{1}{3}} \right].
\end{equation}
Using these two formulas we obtain the isothermal compressibility in explicit form  
\begin{eqnarray}
	\kappa_T^* & = & \frac{\epsilon}{\rho^{*2} W(0)} \left\{   \frac{Q_t^{-1/2}}{\beta W(0) g_4} \left[ \bar \rho_0 - 2 \left( - \frac{3M}{g_4} + \sqrt{Q_t} \right)^{\frac{1}{3}} \right] -1 \right\}.
\end{eqnarray}


\begin{thebibliography}{9}
	
	\bibitem{Johnston2014book} D.C.~Johnston.
	\textit{\href{https://dx.doi.org/10.1088/978-1-627-05532-1}{Advances in Thermodynamics of the van der {W}aals Fluid.}} (Morgan \& Claypool Publishers, 2014) [doi: 10.1088/978-1-627-05532-1]
	
	\bibitem{YigzaweSadus2013} T.M. Yigzawe, R.J.~Sadus.
	\href{https://doi.org/10.1063/1.4803855}{Intermolecular interactions and the thermodynamic properties of supercritical fluids.} \textit{J. Chem. Phys.} \textbf{138}, 194502 (2013) [doi: 10.1063/1.4803855]
	
	\bibitem{Velasco2011} I~Velasco, C.~Rivas, J.F.~Martinez-Lopez, S.T.~Blanco, S.~Otin, M.~Artal.
	\href{https://pubs.acs.org/doi/abs/10.1021/jp202317n}{Accurate values of some thermodynamic properties for carbon dioxide, ethane, propane, and some binary mixtures.} \textit{J. Phys. Chem. B} \textbf{115}, 8216 (2011) [doi: 10.1021/jp202317n] 
	
	\bibitem{Bulavin2024} L.A.~Bulavin, Y.G.~Rudnikov, A.V.~Chalyi.
	\href{https://pubs.aip.org/aip/adv/article/14/8/085213/3307287}{Contributions to the isothermal compressibility coefficient of water near the temperature of 42$^{\circ}$ C.} \textit{AIP Advances} \textbf{14}, 085213 (2024) [doi: 10.1063/5.0205612] 
	
	\bibitem{KozitskyKozlovskiiDobush2018book} Y.~Kozitsky, M.~Kozlovskii, O.~Dobush.
	\href{https://link.springer.com/chapter/10.1007/978-3-319-61109-9_11}{Phase transitions in a continuum Curie-Weiss system: A quantitative analysis.} In: L.~A.~Bulavin, A.~V.~Chalyi (Eds.), \textit{Modern Problems of Molecular Physics} (Springer, 2018) [doi: 10.1007/978-3-319-61109-9\_11]
	
	\bibitem{KozitskyKozlovskiiDobush2020} Y.~Kozitsky, M.~Kozlovskii, O.~Dobush.
	\href{https://icmp.lviv.ua/journal/zbirnyk.102/23502/abstract.html}{A phase transition in a Curie-Weiss system with binary interactions.} \textit{Condens.
		Matter Phys.} \textbf{23}, 23502 (2020) [doi: 10.5488/CMP.23.23502]
	
	\bibitem{PylyukEtAlJML2023}	I.~Pylyuk, M.~Kozlovskii, O.~Dobush, M.~Dufanets.
	\href{https://www.sciencedirect.com/science/article/pii/S0167732223011261}{Morse fluids in the immediate vicinity of the critical point: Calculation of thermodynamic coefficients.} \textit{J. Mol. Liq.} \textbf{385}, 122322 (2023) [doi: 10.1016/j.molliq.2023.122322]
	
	\bibitem{PylyukKozlovskiiDobushUJP2023b} I.~Pylyuk, M.~Kozlovskii, O.~Dobush.
	\href{https://ujp.bitp.kiev.ua/index.php/ujp/article/view/2023197}{Analytic calculation of the critical temperature and estimation of the critical region size for a fluid model.} \textit{Ukr. J. Phys.} \textbf{68}, 601 (2023) [doi: 10.15407/ujpe68.9.601]
	
	\bibitem{KozlovskiiDobush2020} M.~Kozlovskii, O.~Dobush.
	\href{https://ujp.bitp.kiev.ua/index.php/ujp/article/view/2019631}{Phase behavior of a cell fluid model with modified Morse potential.} \textit{Ukr. J. Phys.} \textbf{65}, 428 (2020) [doi: 10.15407/ujpe65.5.428]
	
	\bibitem{PylyukDobush2020} I.~Pylyuk, O.~Dobush. \href{https://ujp.bitp.kiev.ua/index.php/ujp/article/view/2020185}{Equation of state of a cell fluid model with allowance for Gaussian fluctuations of the order parameter.} \textit{Ukr. J. Phys.} \textbf{65}, 1080 (2020) [doi: 10.15407/ujpe65.12.1080]
	
	\bibitem{StrokerMeier2021} P.~Str\"oker, K.~Meier.
	\href{https://link.aps.org/doi/10.1103/PhysRevE.104.014117}{Classical
		statistical mechanics in the grand canonical ensemble.} \textit{Phys. Rev. E} \textbf{104}, 014117 (2021) [doi: 10.1103/PhysRevE.104.014117]
	
	\bibitem{Morse1929}	P.~M.~Morse. \href{https://link.aps.org/doi/10.1103/PhysRev.34.57}{Diatomic
		molecules according to the wave mechanics. II. Vibrational levels.} \textit{Phys.
		Rev.} \textbf{34}, 57 (1929) [doi: 10.1103/PhysRev.34.57]
	
	\bibitem{MartinezValenciaEtAl2013} A.~Mart?nez-Valencia, M.~Gonz?lez-Melchor, P.~Orea, J.~L?pez-Lemus.
	\href{https://doi.org/10.1080/08927022.2012.702422}{Liquid–vapour interface
		varying the softness and range of the interaction potential.} \textit{Molecular
		Simulation} \textbf{39}, 64 (2013) [doi: 10.1080/08927022.2012.702422]
	
	\bibitem{BiswasHamann1985} R.~Biswas, D.R.~Hamann.
	\href{https://link.aps.org/doi/10.1103/PhysRevLett.55.2001}{Interatomic
		potentials for silicon structural energies.} \textit{Phys. Rev. Lett.} \textbf{55}, 2001 (1985) [doi: 10.1103/PhysRevLett.55.2001]
	
	\bibitem{Lim2005} T.-C.~Lim. \href{https://acta-arhiv.chem-soc.si/52/52-2-149.htm}{Approximate
		relationships between the generalized Morse and the extended-Rydberg potential energy functions.} \textit{Acta Chim. Slov.} \textbf{52}, 149 (2005)
	
	\bibitem{KozlovskiiDobush2016} M.~Kozlovskii, O.~Dobush.
	\href{https://www.sciencedirect.com/science/article/pii/S0167732215312101}{Representation of the grand partition function of the cell model: The state equation in the mean-field approximation.} \textit{J. Mol. Liq.} \textbf{215}, 58 (2016) [doi: j.molliq.2015.12.018]
	
	\bibitem{HansenMcDonald2013} J.~Hansen, I.~McDonald. \textit{Theory of Simple Liquids: with Applications to Soft Matter.} 4th Edition (Academic Press, 2013) [ISBN: 9780123870339]
	
	
\end{thebibliography}
\end{document}